# Analysis of atomic magnetometry using metasurface optics for balanced polarimetry

Xuting Yang[1,2], Meryem Benelajla[1,2], and Jennifer T. Choy[1]

[1]University of Wisconsin – Madison, Madison, Wisconsin 53706 USA

[2]These authors contributed equally to this work.

Please send correspondences to jennifer.choy@wisc.edu

**Abstract**

Atomic magnetometry is one of the most sensitive field-measurement techniques for biological, geo-surveying, and navigation applications. An essential process in atomic magnetometry is measurement of optical polarization rotation of a near-resonant beam due to its interaction with atomic spins under an external magnetic field. In this work, we present the design and analysis of a silicon-metasurface-based polarization beam splitter that have been tailored for operation in a rubidium magnetometer. The metasurface polarization beam splitter operates at a wavelength of 795 nm and has a transmission efficiency > 83% and a polarization extinction ratio > 20 dB. We show that these performance specifications are compatible with magnetometer operation in miniaturized vapor cells with subpicotesla-level sensitivity and discuss the prospect of realizing compact, high-sensitivity atomic magnetometers with nanophotonic component integration.

## 1. Introduction

Atomic magnetometers based on optical pumping can provide sub-femtotesla-level sensitivity [1] and are therefore attractive for a host of magnetometry applications including biological sensing [2,3], geo-surveying [4], and magnetic map-based navigation [4,5]. Optically pumped magnetometers (OPMs) typically involve optical pumping of alkali atoms (most commonly rubidium or cesium) with a circularly polarized beam to modify the atomic-spin-dependent optical properties of a medium inside a vapor cell. A linearly polarized probe beam is then used to detect the precession of the atomic spin in the presence of a magnetic field [6]. Typically, this precession causes a polarization rotation of the probe beam which is detected with an optical polarimeter involving a balanced polarimetry, photoelastic or Faraday modulation method [7]. Most OPMs have been realized using an orthogonal pump-probe arrangement with balanced optical polarimetry, since it has a simple configuration and if sufficiently balanced, can suppress common-mode noise arising from laser intensity fluctuations [8].

While breakthroughs in microfabricated vapor cells [9] have enabled the miniaturization of OPMs and their integration with chip-scale photonic technologies such as VCSELs [2,10], state-of-the-art OPMs based on balanced polarimetry detection mostly utilize bulk birefringent polarization optics such as waveplates and polarizing beam splitters which still limit the sensor volume, scalability, and their use in many portable applications [11–13]. More recently, various schemes have been proposed [14,15] and experimentally demonstrated [16,17] to further integrate OPMs with nanophotonic components, but the performance limits of nanophotonic-integrated OPMs are still being explored.

A common approach to simplify OPM designs and make them more scalable towards realizing multi-channel OPM arrays is to have the pump and probe beams share a single optical axis. These inline OPM schemes can be achieved by using a single elliptically polarized beam [18] or by overlapping the pump and probe beams at different transitions (e.g., the D1 and D2 lines) [19]. Inline OPM configurations are also highly compatible with miniaturized magnetometers since microfabricated vapor cells usually allow only one optical access path.

In this work, we propose integration of inline OPMs with metasurface-based polarization components (Fig. 1). Specifically, we have designed an efficient and compact balanced polarimetry scheme using a metasurface-based polarization beam splitter (PBS). The geometry of our metasurface-based PBS can be freely optimized and tailored for integration with other atomic species and single-beam OPM designs. In addition, with advances in MEMS vapor cell fabrication [20] as well as techniques to integrate atomic vapor with integrated photonic circuits [16,17,21–23], it should be possible to directly fabricate metasurfaces on the glass windows of atomic-vapor cells.

We then evaluate the accuracy and sensitivity of atomic magnetometry using the miniaturized balanced detection scheme. For our OPM platform, we use Spin Exchange Relaxation Free (SERF) magnetometry in rubidium (Rb) with elliptically polarized light [16]. SERF magnetometers typically operates in a high atomic density regime ($>10^{13}\ cm^{-3}$) where the spin-exchange collision rate of the alkali metal atoms is much larger than the Larmor precession frequency, resulting in longer ground state Zeeman coherence time and thus extremely high sensitivity [18]. In this case, the circularly polarized component of the elliptically polarized light induces atomic polarization in the ground state of isotopically purified Rb atoms through optical pumping. A balanced polarimetry scheme is then used to measure the optical rotation of the linearly polarized component of the transmitted light at the output of the cell. The balanced polarimeter consists of a metasurface PBS that splits the horizontal and vertical linear polarization components of the light beam propagating through the vapor cell medium which are then measured using two independent photodetectors. The detection of the difference of the horizontal and vertical linear polarization intensities provides a direct measure of the optical rotation. The sensitivity of the magnetometer is determined by the signal-to-noise ratio with which the rotation signal is measured.

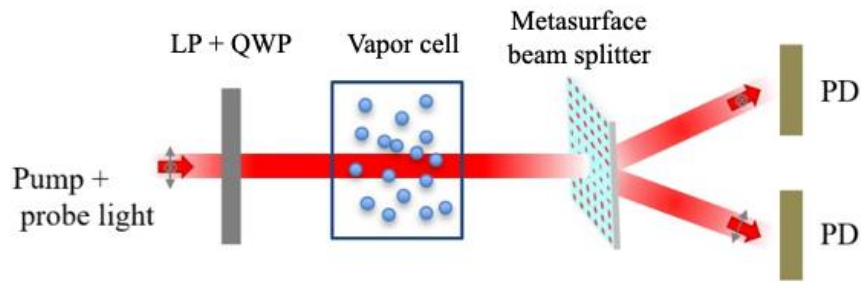

FIG. 1. Schematic of the proposed atomic magnetometer with a metasurface-based polarimeter. LP: linear polarizer; QWP: quarter waveplate; PD: photodetector.

The paper is structured as follows. In section 2, we introduce our approach to design nanophotonic components for miniaturized balanced detection scheme in in-line OPMs. Optical performance of those components is discussed including transmission losses and polarization extinction ratio. A performance analysis of SERF with metasurface optics is presented in section 3.

## 2. Design and modeling of metasurface-based PBS

*2.1 Design approach*

Recently, dielectric metasurfaces made of arrays of elliptical posts have been adopted to design a wide range of nanophotonic components (such as beam splitters, waveplates and optical lenses) offering efficient control of the amplitude, phase, and polarization of light. The majority efforts have been dedicated to engineer metalenses [25], beam shaping devices [26,27], polarization optics [28] and holograms [29]. These advances have recently been applied to trap and cool atoms on a chip [30]. In this paper, we adapt an approach to design metasurface polarization components for operation at near-infrared (780 - 795 nm) wavelengths and thus compatible with Rb-based sensors. Our designs maintain high-transmission efficiency and polarization extinction ratio. Our approach is based on the formalism described by Arbabi et al in [28]. Specifically, we implement metasurface optical components by arrays of elliptical posts with the same height, but different diameters to locally modify the phase distribution and polarization of any arbitrary input beam.

We consider a transmissive metasurface polarizing beam splitter design (Fig. 2a) that can be used in a variety of miniaturized OPMs for balanced polarimetry of the magnetometry signal resulting from the polarization rotation of the probe light. For such a metasurface design, the optical response is fully determined by the phase and amplitude of light in transmission. Fig. 2b shows a schematic illustration of the unit cell structure of the metasurface, consisting of a single-crystalline silicon elliptical post of height $H$ and post diameters $r_x$ and $r_y$ along the $x$ and $y$ axes on a sapphire substrate. In the design, $H$ is fixed while $r_x$ and $r_y$ are tailored to achieve the desired $\varphi_x$ and $\varphi_y$ phase shifts for $x$ and $y$-polarized waves, respectively. To implement beam splitting based on input polarization of a normally incident beam, the desired phase shifts are set to $\varphi_x = -kx\sin\theta$ for an $x$-polarized beam and $\varphi_y = kx\sin\theta$ for a $y$-polarized beam, where the wavenumber is $k = 2\pi/\lambda$ and $2\theta$ is the separation angle between the split beams. For this design, the phase shifts are invariant along $y$. We note that the metasurface splits the incident light into beams with polarizations along $xz$ (from the $x$-polarized component of the incident beam) and $y$ (from the $y$-polarized component of the incident beam). For simplicity, we will refer the $xz$-polarized transmitted beam as the $x$-polarized beam for the remainder of the paper.

Next, we construct the metasurface by sampling the phase profile using a square lattice of silicon elliptical posts of various diameters that implement the required phase shift at that position. As stated in [28], to avoid non-zero order diffraction and to achieve a better approximation of the phase profile, the sampling lattice constant $a$ should be smaller than the operating wavelength. In our case, we choose $a = 400\ nm$ for the Rb D1-line ($795\ nm$).

By varying the geometric parameters $r_x, r_y$ of the elliptical posts, we can impose independent phase shifts along the $x$ and $y$ polarization axes. Figs. 2c and d show the simulated transmitted power and phase shift of an array of elliptical posts with different axis radii from 40 nm to 140 nm for an incident linearly polarized ($x$ or $y$) plane wave. We see directly from the figure that this platform provides a complete phase coverage independently over $\varphi_x$ and $\varphi_y$ while maintaining a relatively high transmission amplitude of $> 87\%$. Elliptical posts of specific radius are selected to match the metasurface phase profile at their position by minimizing the squared error $\epsilon = |t_x - e^{i\varphi_x}|^2 + |t_y - e^{i\varphi_y}|^2$, where $t_i$ is the transmission coefficient of the elliptical post and $\varphi_i$ is the calculated phase profile for polarization $i = x, y$, respectively. The resulting metasurface consists of $2500 \times 2500$ posts (a portion of which is shown in Fig. 2e) and is designed to achieve a $2\theta = 20°$ split angle between the $x$ and $y$ polarization components. For this design, the maximum $\epsilon$ obtained is 0.0886 while the average $\epsilon$ is 0.0166. These $\epsilon$ values are limited by the mismatch between the desired phase and amplitude profiles and the discretized values in Figs. 2c and 2d, but are nonetheless sufficiently low to ensure good beam splitting performance while maintaining high transmission, as will be shown in the next section.

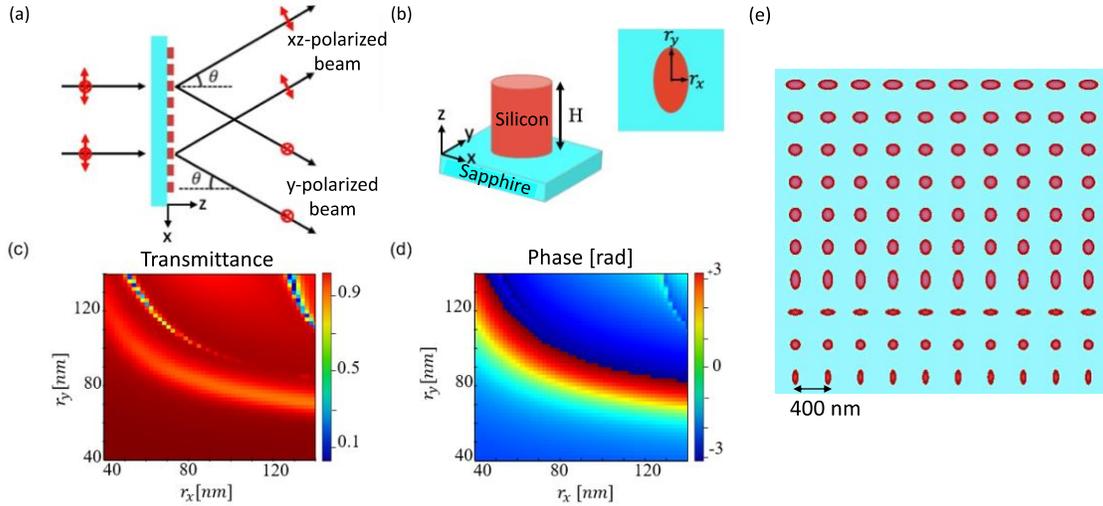

FIG. 2. (a) Operation of a metasurface polarizing beam splitter, which splits an incident beam into beams of orthogonal xz and y polarizations. Note that for simplicity, we will refer the xz-polarized beam as the x-polarized transmitted beam for the rest of the paper. (b) Illustration of a unit cell comprising of a silicon post on a transparent substrate (here, sapphire). The silicon post has an elliptical cross-section (principal axis lengths $r_x$ and $r_y$) and height H. (c)-(d) Calculated normalized transmitted power ("Transmittance") and phase shift of an incident x- or y- polarized plane wave, based on full-wave simulation of swept elliptical parameters with fixed lattice constant (a = 400 nm) and post height (H = 500 nm). (e) Top view of a section of the metasurface polarization beam splitter design.

*2.2 Simulated optical performance*

To characterize the optical performance of our metasurface beam splitter, we performed finite-difference time domain (FDTD) simulations for a design with a device area of 16 μm × 16 μm using a Gaussian beam source (see Supplementary Information S1 on justification of our simulated beam source). Our design can be extended to a much larger (>1 mm$^2$) device footprint, but this representative simulation area is chosen to limit computational time. The incident beam of wavelength $\lambda = 795\ nm$ (D1 transition of $^{87}Rb$) is propagating in the $z$-direction at normal incidence. We evaluated the transmission characteristics of the metasurface for different input polarization angles. Fig. 3a represents the normalized transmitted far-field intensity for an incident linear polarization at 45° (comprising of an equal superposition of $x$ and $y$ polarizations). As expected, the designed metasurface splits the 45° linearly polarized incident beam into $x$- and $y$-polarized beams propagated along different directions. The split beams have a slightly different transmittance with a split ratio of 49% and 47% for $x$ and $y$ polarization respectively. The transmittance among various input polarization angles is shown in Fig. 3b. The results for $x$-polarized beam (respectively for $y$-polarized beam) demonstrate that the transmitted intensity varies as the square of the cosine (respectively sine) of the polarization angle of the input beam. This is in good consistency with Malus law for linear polarizers.

The performance shown here is for input beams of wavelength $\lambda = 795\ nm$ at normal incidence, in which the simulated transmittance is over 80% exhibits a slight linear dichroism (83.3% for $x$- polarized beam and 85.8% for $y$-polarized beam). Polarization extinction ratio (PER) is another important performance metric for PBS devices. We define the PER for each output port as the light intensity of main polarization mode divided by the light intensity of the orthogonal polarization mode (see Supplementary Information S2). For normal incidence, the PER is 20.55 dB for $x$ polarization and 28.3 dB for $y$ polarization. The PER values of this design are limited by the discrete sampling of the metasurface and the residual phase error ($\epsilon$ in Section 2.1) when matching the geometric parameters of each elliptical post with the desired phase response.

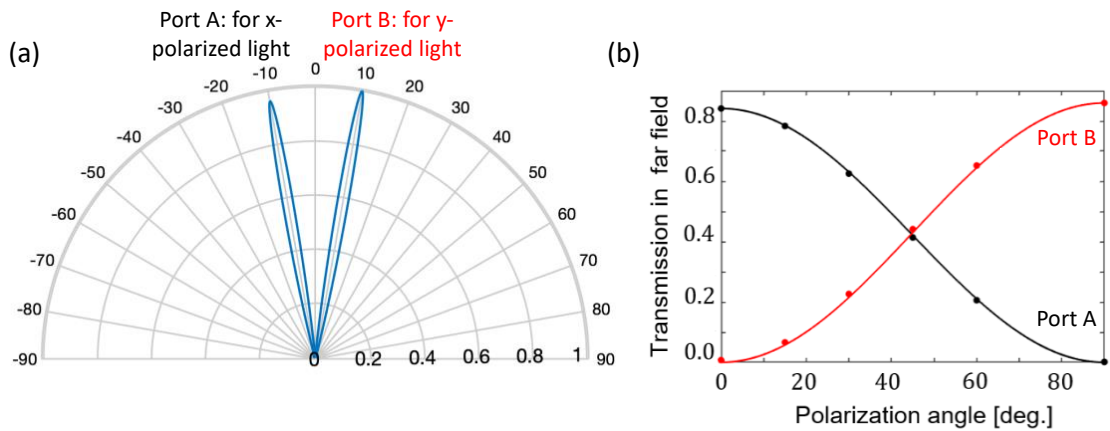

FIG. 3. (a) The 3D far-field normalized intensity scattering patterns for the 45-linearly polarized incident Gaussian beam under normal incidence. (b) The normalized transmission in far field for different polarization angles of the input beam. Each data point for the $x$ and $y$-polarized beam is obtained by integrating the optical power within their respective scattered spot, then normalize it to the overall transmission power. The solid lines indicate sinusoidal fit to the data.

In reality, the transmission efficiency and PER of the metasurface structure may vary with the incident angle as it has been reported in earlier works [15]. These deviations are significant considerations in an OPM, which may have axial misalignments between an input beam and the atomic cell. We have analyzed the transmission and polarization extinction of our design for different angles of incidence to evaluate the alignment requirements for achieving sensitive balanced polarimetry measurements. Figs. 4c and 4d show the transmittance and polarization extinction ratio at different incident angles from 0 to 10°. Both performance metrics suffer from degradation as the incident angle exceeds 5°, which is again attributed to the higher-order diffraction as the incidence condition deviates from the requirement for zero-order diffraction: $\lambda \gg \frac{a}{\cos\theta}$ (where $\theta$ is the incident angle). Nevertheless, the polarization extinction ratio maintains a level of $> 17\ dB$ over a range of 10°, an alignment condition that can be readily achieved in the actual experiment.

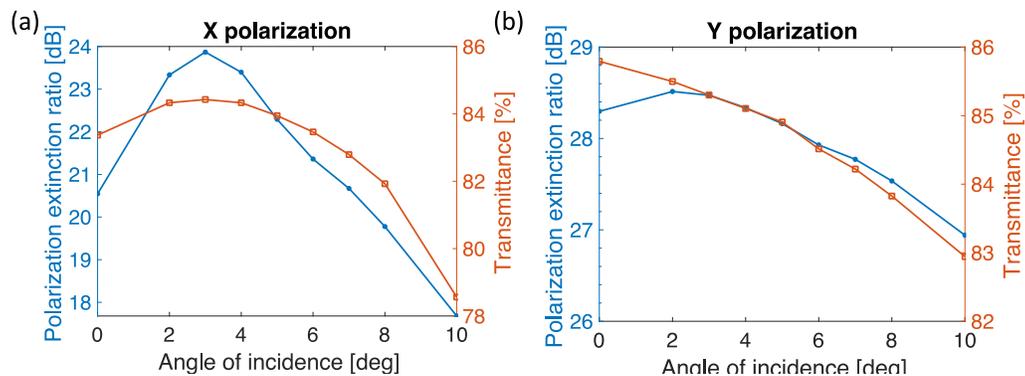

FIG. 4. Simulated intensity transmission and polarization extinction ratio as function of the incident angle for (a) $x$-polarized beam and (d) $y$-polarized beam.

## 3. Sensitivity performance analysis

In this section, we evaluate how the optical performance of our proposed metasurface PBS affects balanced polarimetry and thereby the sensitivity and accuracy of an OPM. Our PBS design can is compatible with millimeter-scale, microfabricated Rb cells involving separate non-parallel [15] or inline pump and probe beams. For reasons specified in Section 1, we focus on the magnetometry schemes that are compatible with inline pump-probe configurations such as the Mx [31], Bell-Bloom [32], SERF [33], and nonlinear magneto-optical rotation (NMOR) [34] OPMs. Of the above schemes proposed, SERF is an attractive candidate since it provides the highest sensitivities among OPMs and has been demonstrated to achieve sub-picotesla sensitivity in microfabricated cells [33]. Moreover, while the shot-noise-limited sensitivity of SERF magnetometry is expected to degrade with decreasing cell size [35], SERF operation relies on a high buffer gas pressure which is also necessary in very small ($\lesssim$ mm$^3$) vapor cells to prevent decoherence from collisions of the atoms with the cell wall. A high buffer gas pressure would broaden atomic transitions and preclude the operation of NMOR [36]. Finally, SERF using elliptically polarized light has shown very promising results with device scale down to 5 mm [16, 31] and can require only a single optical beam to operate [18,19], and thus highly compatible with nanophotonic integration.

We thus choose SERF magnetometry using a single elliptical beam as the platform for evaluating the sensitivity of a metausrface-integrated magnetometer [18]. In this scheme, an elliptically polarized light close to the $^{87}Rb$ D1 transition interacts with a $^{87}Rb$ vapor cell of optical path length $l$ and Rb number density $n$ to generate atomic spin polarization. This same optical beam is used to detect spin polarization via balanced polarimetry. An amplitude-modulating field is applied along the direction of magnetic field measurement (along $x$) which must be transverse to the light propagation direction; all other field components are nulled. The optical signal is then detected using lock-in detection at the modulation frequency ($\omega_{mod}$). The theory of operation for SERF magnetometry with elliptical light was worked out in [18] for the case of an ideal PBS with transmittance $T = 1$ and infinite polarization extinction ratio. In this scenario, the balanced polarimetry gives the polarization angle rotation $\phi$ via the differential signal of the two outputs from the PBS:

$$\mathcal{D} = E_0^2 e^{-\sigma n l} \cos 2\beta \sin\phi \qquad (1)$$

where $\phi$ is proportional to the spin polarization along the optical axis (see Supplementary Information S3 for derivation); $E_0$ is the electric field amplitude; $e^{-\sigma n l}$ is related to light absorption by Rb atoms (whereby $\sigma$ is the photon absorption cross section for unpolarized Rb atoms); $\beta$ is the ellipticity of the light and is set to $\pi/8$ [18]. In the low polarization limit, the signal $\mathcal{D}$ is linear in terms of spin polarization $P_z$. This optical rotation will allow us to determine the transverse magnetic field.

*3.1 Effects of polarization-dependent transmittance and PER on magnetometry*

Note that the differential signal $\mathcal{D}$ shown in Eq. 1 only includes the polarization rotation term induced by circular birefringence. Such simplicity cannot be assumed for our metasurface PBS which has a non-negligible transmittance difference for the two orthogonal linear polarization and polarization leakage. We thus modify the theory to incorporate these non-idealities. Let $T_x, T_y$ be the transmittance for the $x$ and $y$ polarizations, $b_x$ and $b_y$ be the percentage of the transmitted optical power that is leaked into the other output of the PBS for a pure $x$ and $y$ polarization input, respectively. $b$ is related to PER through $b = \frac{1}{PER+1}$. Then the outputs of the PBS become:

$$\begin{aligned} I_x &= T_x(1-b_x)I_{x,in} + T_y b_y I_{y,in} \\ I_y &= T_x b_x I_{x,in} + T_y(1-b_y)I_{y,in} \end{aligned} \quad (2)$$

i.e. the two outputs are now a linear superposition of the light intensities along two orthogonal polarization axes. The direct consequence of this intensity mixture is that the differential signal will now contain an additional term due to circular dichroism that is proportional to $\sigma$. It can be shown that under this "imperfect PBS" model, the differential signal can be approximated as (see Supplementary Information S3 for the derivation):

$$\mathcal{D} = E_0^2 e^{-\sigma n l}(\eta \cos 2\beta \sin\phi + \frac{\left(T_y(1-2b_y) - T_x(1-2b_x)\right)}{4} \sigma n l \sin 2\beta P_z) \quad (3)$$

where

$$\eta = \frac{T_x(1-2b_x) + T_y(1-2b_y)}{2}\sqrt{1 - \left(\frac{T_y(1-2b_y) - T_x(1-2b_x)}{\cos 2\beta(T_x(1-2b_x) + T_y(1-2b_y))}\right)^2}; \quad (4)$$

$\phi$ is the polarization rotation angle and is proportional to $P_z$. We see that the imperfect PBS results in two modifications in the expression of the differential signal: first, the original polarization rotation signal is attenuated by a factor of $\eta$; second, the imperfect PBS introduces another term related to the change of the polarization ellipticity induced by the circular dichroism of the atomic medium, which is also proportional to $P_z$ in the low polarization limit.

We will now compare the magnetometry signal of a metasurface-integrated magnetometer with one based on ideal optics (namely, a PBS with near unity transmittance, ultra-high PER, and no linear dichroism). We use identical excitation and cell conditions as in [18], in which the Rb vapor cell is filled with 300 Torr helium and 100 Torr nitrogen and heated to 200 ℃ . The laser frequency detuning is set to $v_0 - v = 45\ GHz$. These conditions lead to a linewidth $\Delta v$ of 7.97 GHz, optical pumping rate of 880 Hz, and a spin relaxation rate of 1200 Hz. Fig. 5 shows the differential signal $\mathcal{D}$ as a function of the transverse magnetic field $B_x$ for magnetometers based on ideal optics (red) and our metasurface design (blue) for a vapor cell with $l = 5\ mm$. We can see from the Fig. 5 that the incorporation of the metasurface PBS into the magnetometer does not change the dispersive character of the signal and only modifies the signal amplitude by a scaling factor $\xi$ that is given by:

$$\xi = \eta + \frac{\left(T_y(1-2b_y) - T_x(1-2b_x)\right)}{8} \frac{\Delta v}{v_0 - v} \qquad (5)$$

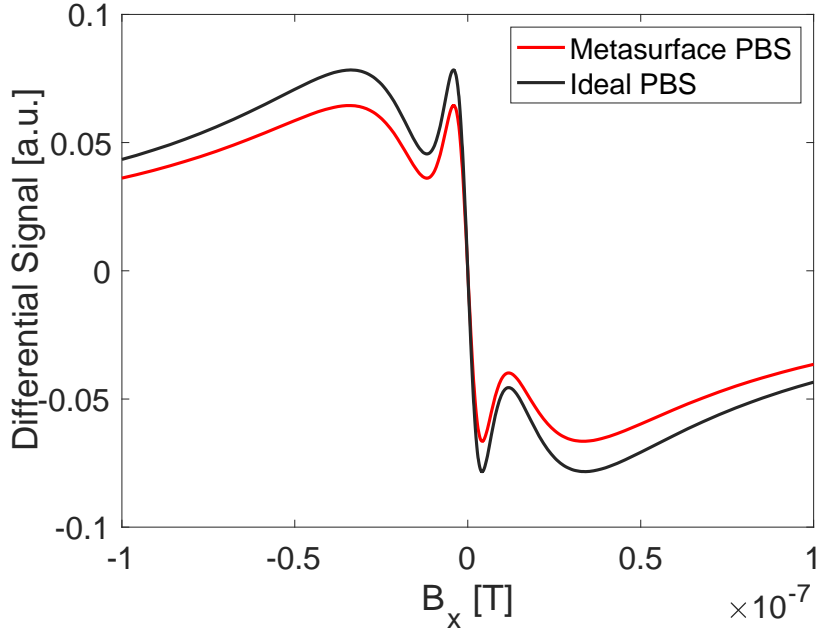

FIG. 5. Simulated differential signal $\mathcal{D}$ with lock-in detection at the modulating frequency $\omega_{mod}$ as a function of $B_x$ with an ideal PBS (red) and metasurface PBS (blue).

For our design, $\xi \cong 0.837$. The metasurface polarization extinction ratios are still sufficiently high, such that $b_x, b_y \ll 1$. Additionally, there is with no significant linear dichroism $|T_x - T_y| \ll 1$ and the laser detuning is sufficiently larger than the pressure-broadened linewidth. Under these conditions, the signal amplitude is simply attenuated by the mean transmittance of $\hat{x}$ and $\hat{y}$ polarization: $\xi \cong \frac{T_x + T_y}{2}$. Finally, it may be possible to further improve the metasurface design using adjoint optimization methods [37–39] that minimize the figure of merit (FOM) defined as $FOM = \left|\frac{T_y(1-2b_y)-T_x(1-2b_x)}{T_x(1-2b_x)+T_y(1-2b_y)}\right|$. This analysis can also serve as a useful guideline for designing nanophotonic components for other quantum sensors.

*3.2 Impact of metasurface PBS on magnetometer noise*

We now discuss the impact of the metasurface PBS on the various noise sources accompanying the atomic magnetometer signal. It is shown in [33] that in a single-beam, millimeter-scale magnetometer based on absorption measurement, the laser intensity noise dominates. The laser intensity noise can arise from laser current fluctuations or light polarization drifts that are converted into amplitude noise by a linear polarizer at the input. The SERF scheme we adopted for miniaturization in this paper uses balanced polarimetry as well as spin modulation with lock-in detection, which are capable of suppressing the intensity noise [18,40]. A sensitivity of $7\ fT/\sqrt{Hz}$ was reported in a centimeter-scale cell in [18], with the dominant noise source being the Johnson noise from the magnetic shield.

The residual laser intensity noise will contribute to the overall technical noise (which also includes electronic noise) which will limit the signal-to-noise ratio (SNR) of the measurement. Since the metasurface PBS scales the differential signal amplitude by $\xi \cong 0.837$, we expect the SNR of the metasurface-integrated magnetometer to scale by a similar amount. Based on the inverse relationship between sensitivity and SNR [41], the magnetic field sensitivity will thus degrade by about a factor of 1.2.

In addition to its impact on detection, laser intensity noise also affects the optical pumping rate and light shifts experienced by atoms [42,43], We calculated the effect of optical pumping rate variation in Supplementary Information S4 and showed that the resulting field measurement errors are not affected by the use of a metasurface PBS. The light shift effect is equivalent to a fictitious magnetic field along the light propagation direction ($z$) with its amplitude proportional to the laser intensity. In [18], the light shift effect is countered by applying a magnetic field in the opposite direction that minimizes the magnetic resonance linewidth. If the laser intensity noise is present, there will always be a small fictious field $\delta B_z$ that is not balanced. Fortunately, if the noise amplitude is small enough such that $\gamma \delta B_z \tau \ll 1$, then it is shown in [44] that the induced change in atomic polarization $P_z$ is of second order in $\delta B_z$, rendering the error negligible.

**4. Conclusion and outlook**

We have demonstrated a metasurface-based balanced polarimeter design for a compact Rb magnetometer based on SERF and have connected its optical performance to the magnetometer accuracy and sensitivity. This design can be realized with electron-beam lithography and reactive-ion etching techniques on single-crystal silicon films grown on transparent substrates. It may be possible to incorporate the metasurface PBS as part of the glass wafer anodically bonded to the silicon housing during MEMS fabrication of mm-scale vapor cells [10]. Modeling of balanced polarimetry using simulated transmittance and PER shows that the nanophotonic component can be integrated into SERF magnetometry with only a 20% degradation in sensitivity. Our analysis of the magnetometry dependence of an imperfect PBS may inform future efforts to design and integrate nanophotonic polarization optics into atomic magnetometers and other atom-based sensors.

**Acknowledgments**

This work is supported by the Office of Naval Research under Grant No. N00014-20-1-2598 and the Wisconsin Alumni Research Foundation. The authors thank Mikhail Kats and Thad Walker for helpful discussions on the metasurface design and its suitability to atomic magnetometry platforms. We also thank Jinsheng Hu for a careful reading of our manuscript and providing valuable feedback on our analysis.

**Disclosures** The authors declare no conflicts of interest.

**Data availability** Data underlying the results presented in this paper are not publicly available at this time but may be obtained from the authors upon request.

**Supplemental document** See Supplementary Information.

Supplementary Information for "Analysis of atomic magnetometry using metasurface optics for balanced polarimetry"

Xuting Yang[1,2], Meryem Benelajla[1,2], and Jennifer T. Choy[1,*]

[1]University of Wisconsin – Madison, Madison, Wisconsin 53706 USA

[2]These authors contributed equally to this work.

## S1: Plane-wave approximation

Here we justify our assumption that the phase profile $\varphi(x, y)$ of our metasurface polarizing beamsplitter that is designed for a normal plane wave incidence would still apply to the case where the incident beam has a Gaussian profile. Let us consider a Gaussian source polarized along the $x$ direction:

$$U(x,y,z) = E_0 \hat{x} \frac{w_0}{w(z)} \exp\left(-\frac{x^2+y^2}{w(z)^2}\right) \exp\left(-i\left(kz + k\frac{x^2+y^2}{2R(z)} - \psi(z)\right)\right) \quad \text{(Eq. S1)}$$

where $w_0$ is the waist radius, $w(z) = w_0\sqrt{1+(\frac{\lambda z}{\pi w_0^2})^2}$ is the radius at which the field amplitudes decay to $\frac{1}{e}$ of their axial values, $R(z) = \frac{z^2 + (\frac{\pi w_0^2}{\lambda})^2}{z}$ is the radius of curvature, $\psi(z) = \arctan(\frac{\lambda z}{\pi w_0^2})$ is the Gouy phase.

Our main argument here is that $U(x, y, z)$ *is* a solution to the paraxial Helmholtz equation, and can thus be well-approximated by a normal incident plane wave plus a small component of paraxial wave:

$$\begin{aligned} U(x,y,z) &= \iint u(k_x, k_y) e^{-i(k_x x + k_y y + k_z z)} dk_x dk_y \\ &= \iint e^{-ikz} u(k_x, k_y) e^{-i\left(k_x x + k_y y - \frac{1}{2k}(k_x^2 + k_y^2) z\right)} dk_x dk_y \end{aligned} \quad \text{(Eq. S2)}$$

where we have used the fact that $k_x, k_y \ll k_z$.

We claim that the elliptical posts will impose the same phase shift $\varphi(x,y) = \mp kx\sin\Theta$ for a paraxial plane wave, so that the gaussian source after transmitting through the metasurface will become: $U'(x,y,z) = \iint e^{-ik(z \mp x\sin\Theta)} u(k_x, k_y) e^{-i\left(k_x x + k_y y - \frac{1}{2k}(k_x^2 + k_y^2)z\right)} dk_x dk_y$ which is just the gaussian beam deflected by angle $\mp\Theta$. In what follows, we justify our claim. Let us quantify this argument by calculating the Fourier component $u(\theta)$ of a Gaussian beam, where $\theta$ indicates the angle between the wavevector and the propagation direction $\hat{z}$. We write the k-vector in spherical basis: $k_x = K\sin\theta\cos\varphi$, $k_y = K\sin\theta\sin\varphi$, $k_x = K\cos\theta$. Here $\varphi$ is the azimuth angle.

$$u(K, \theta) = \iiint U(x,y,z) e^{i(k_x x + k_y y + k_z z)} dx\, dy\, dz \quad \text{(Eq. S3)}$$

$$= E_0 w_0 \int_{-\infty}^{+\infty} \frac{1}{w(z)} e^{i(K\cos\theta - k)z + \psi(z))} dz$$

$$\times \iint e^{-\frac{x^2+y^2}{w(z)^2}} e^{i(K\sin\theta\sin\varphi x + K\sin\theta\sin\varphi y - k\frac{x^2+y^2}{2R(z)})} dxdy$$

It can be proven that

$$|u(K,\theta)| = C(w_0, \lambda, E_0)|g(z,)|e^{-\frac{K^2 w_0^2}{4}\sin^2\theta} \quad \text{(Eq. S4)}$$

Here $C(w_0, \lambda, E_0)$ is a constant that only depends on the incident Gaussian beam parameters. $g = \int_{-\infty}^{+\infty} \frac{e^{i(K\cos\theta-k)z+\psi(z)}}{\sqrt{1+(\frac{\lambda z}{\pi w_0^2})^2}} dz$ is a complex function of $\theta$.

If $\alpha = K\cos\theta - k \neq 0$, then $g(\alpha) = \int_{-\infty}^{+\infty} \frac{e^{i(\alpha z + \psi(z))}}{\sqrt{1+(\frac{\lambda z}{\pi w_0^2})^2}} dz$ converges, therefore the Fourier components that matter are the ones with $K = \frac{k}{\cos\theta}$. We now have:

$$|u(\theta)| \propto e^{-\frac{k^2 w_0^2}{4}\tan^2\theta} \quad \text{(Eq. S5)}$$

The angular dependence of the Fourier component decays very fast with respect to $\theta$. We define a cutoff angle $\theta_0$ where its corresponding Fourier coefficient is only $e^{-4} \approx 1.8\%$ of that of the normal component. Then:

$$\boldsymbol{\theta_0 = arctan\left(\frac{4}{kw_0}\right)} \quad \text{(Eq. S6)}$$

For our simulation, $w_0 = 4\ \mu m$ and $k = 7.9 \times 10^6 m^{-1}$, the cutoff angle $\theta_0 = 7°$. For a realistic beam size $w_0 = 1\ mm$ in the magnetometer experiment, $\theta_0 = 0.03°$. It follows that the Gaussian source in the actual experiment is very well approximated by a normal incident plane wave. Indeed, simulated phase shift of a 795-nm plane wave with incident angle at the cutoff angle $\theta_0 = 0.03°$ shows less than $10^{-6}$ deviation from that for a normally incident plane wave.

### S2: Determination of transmittance and polarization extinction ratio

In this section we describe our method for calculating the transmittance $T$ and the polarization extinction ratio (PER) of the metasurface PBS. The transmittance $T_0$ from the transmission of the field through the metasurface is first extracted from a field monitor placed a few wavelengths away from the metasurface. In the far-field projection, the transmitted field intensity $|E|^2$ is integrated at the two PBS outputs whose locations are determined by the splitting angle $2\theta$. The integration area is determined through the solid angle $\Omega$ of the projection hemisphere. We choose $\Omega = 2\pi(1 - \cos 7°)$ (half angle $7°$) to ensure we have covered most of the scattered light and the two integration areas do not overlap. The reported transmittance values in the main text take into account both the initial transmittance $T_0$ and the fraction of intensity within the solid angle:

$$T = T_0 \times \frac{\int_{cone} |E|^2}{\int_{hemisphere} |E|^2} \qquad (\text{Eq. S7})$$

We now define $I_{main}$ to be the integrated intensity in the far field that corresponds to the main polarization mode and $I_{leak}$ to be the intensity corresponding to the orthogonal polarization mode, then the PER is calculated as:

$$PER = \frac{I_{main}}{I_{leak}} \qquad (\text{Eq. S8})$$

## S3: Derivation of differential signal for an imperfect PBS

In this section we show the derivation of the differential signal between the output ports of an imperfect PBS, for SERF magnetometry with a single elliptical beam [1]. We write an electric field with arbitrary polarization and amplitude in the $x$ and $y$ polarization basis: $E = c_1 \hat{x} + c_2 i \hat{y}$ where $c_1$ and $c_2$ are complex numbers.

If the optical axes of the PBS are oriented with respect to the $x - y$ coordinate by angle $\delta$, then the two outputs of the metasurface PBS are:

$$\begin{aligned} I_1 &= \mathcal{P}\big(T_x(1-b_x)|c_1 \cos\delta + i c_2 \sin\delta|^2 + T_y b_y |-c_1 \sin\delta + i c_2 \cos\delta|^2\big) \\ I_2 &= \mathcal{P}(T_x b_x |c_1 \cos\delta + i c_2 \sin\delta|^2 + T_y(1-b_y)|-c_1 \sin\delta + i c_2 \cos\delta|^2) \end{aligned} \qquad (\text{Eq. S9})$$

Here $\mathcal{P}$ is some constant is related to vacuum permittivity and detection efficiency. With the differential signal being:

$$\begin{aligned} \mathcal{D} &= I_1 - I_2 \\ &= \mathcal{P}[T_x(1-2b_x)|c_1\cos\delta + ic_2\sin\delta|^2 - T_y(1-2b_y)|-c_1\sin\delta + ic_2\cos\delta|^2] \\ &= \mathcal{P}[T_x(1-2b_x)(|c_1|^2\cos^2\delta + i(\bar{c_1}c_2 - c_1\bar{c_2})\sin\delta\cos\delta + |c_2|^2\sin^2\delta) \\ &\quad - T_y(1-2b_y)(|c_1|^2\sin^2\delta + i(c_1\bar{c_2} - \bar{c_1}c_2)\sin\delta\cos\delta \\ &\quad + |c_2|^2\cos^2\delta)] \end{aligned} \qquad (\text{Eq. S10})$$

In the absence of an atomic medium, $|c_1|^2 = \cos^2\beta, |c_3|^2 = \sin^2\beta, \bar{c_1}c_2 - c_1\bar{c_2} = 0$. Let $\mathcal{D} = 0$ and we get the expression for $\delta$ that balances the two outputs:

$$\tan^2\delta = \frac{T_x(1-2b_x)\cos^2\beta - T_y(1-2b_y)\sin^2\beta}{T_y(1-2b_y)\cos^2\beta - T_x(1-2b_x)\sin^2\beta} \qquad (\text{Eq. S11})$$

The interaction of the field with rubidium atoms causes polarization rotation and ellipticity change, with the coefficients now being:

$$|c_1|^2 = \frac{1}{2}[(e^{-n\sigma l(1+P_z)} + e^{-n\sigma l(1-P_z)}) + (e^{-n\sigma l(1+P_z)} - e^{-n\sigma l(1-P_z)})\sin2\beta \quad \text{(Eq. S12)}$$
$$+ 2e^{-n\sigma l}\cos2\beta \cos\phi]$$

$$|c_2|^2 = \frac{1}{2}[(e^{-n\sigma l(1+P_z)} + e^{-n\sigma l(1-P_z)}) + (e^{-n\sigma l(1+P_z)} - e^{-n\sigma l(1-P_z)})\sin2\beta \quad \text{(Eq. S13)}$$
$$- 2e^{-n\sigma l}\cos2\beta \cos\phi]$$

$$\bar{c}_1 c_2 - c_1 \bar{c}_2 = -2ie^{-n\sigma l}\cos2\beta \sin\phi \quad \text{(Eq. S14)}$$

where $\phi = cr_e fnlP_z Re(L(\nu))$ is the polarization rotation angle defined in [1]. Here $c$ is the speed of light in vacuum; $r_e$ is the classical electron radius; $f \cong \frac{1}{3}$ is the oscillator strength of the rubidium D1 transition; $l$ is the optical path length; $n$ is the Rb number density; $P_z$ is the atomic spin polarization along z direction; $L(\nu) = \frac{1}{\nu_0 - \nu + i\frac{\Delta\nu}{2}}$ is the pressure-broadened Lorentzian profile with a full width at half maximum $\Delta\nu$ centered at $\nu_0$. Plug in the expressions for $c_1, c_2$ and $\delta$ into expression (Eq. S10), and select $\mathcal{P}$ such that our result reduces to the expression given in [1] when $T_x = T_y = 1, b_x = b_y = 0$, we have:

$$\mathcal{D} = \frac{1}{8}E_0^2[(T_x(1-2b_x) - T_y(1-2b_y))$$
$$\times (e^{-n\sigma l(1+P_z)} + e^{-n\sigma l(1-P_z)} + \sin2\beta(e^{-n\sigma l(1+P_z)} - e^{-n\sigma l(1-P_z)}))$$
$$+2(T_y(1-2b_y) - T_x(1-2b_x))e^{-n\sigma l}\cos\phi) \quad \text{(Eq. S15)}$$
$$+4(T_x(1-2b_x) + T_y(1-2b_y))$$
$$\times \sqrt{1 - \left(\frac{T_y(1-2b_y) - T_x(1-2b_x)}{\cos2\beta(T_x(1-2b_x) + T_y(1-2b_y))}\right)^2} e^{-n\sigma l}\cos2\beta \sin\phi)]$$

Expanding and keep the first order of $P_z$, we have:

$$\mathcal{D} = E_0^2 e^{-\sigma nl} \big(\frac{T_x(1-2b_x) + T_y(1-2b_y)}{2}$$
$$\times \sqrt{1 - \left(\frac{T_y(1-2b_y) - T_x(1-2b_x)}{\cos2\beta(T_x(1-2b_x) + T_y(1-2b_y))}\right)^2} \cos2\beta \sin\phi \quad \text{(Eq. S16)}$$
$$+ \frac{(T_y(1-2b_y) - T_x(1-2b_x))}{4}\sigma nl\sin2\beta P_z\big)$$

For $\phi, \sigma nlP_z \ll 1$, $\mathcal{D}$ reduces to

$$\mathcal{D} = \xi E_0^2 e^{-\sigma nl}\cos2\beta \sin\phi \quad \text{(Eq. S17)}$$

where $\xi$ is defined in Eq. 5 in the main text.

Let us now discuss what conditions should the transmittance and PER of the PBS satisfy to generate a dispersive signal with discernible amplitude. Note that under the SERF condition, the atomic vapor is usually dense enough that the modification of the optical absorption due to spin polarization $\Lambda = n\sigma l P_z$ is no longer small enough for the first order approximation to remain valid (except for sufficiently small vapor cells, see Fig. S1). Therefore, the polarimetry signal from circular dichroism (which is the one related to $Im(L(\nu))$) is no longer dispersive due to the even power terms in the expansion. We thus require this term to be suppressed compared to the term due to polarization rotation:

$$\frac{|T_y(1-2b_y) - T_x(1-2b_x)|}{8} \tan 2\beta \frac{\Delta \nu}{\nu_0 - \nu}$$
$$\ll \frac{T_x(1-2b_x) + T_y(1-2b_y)}{2} \sqrt{1 - \left(\frac{T_y(1-2b_y) - T_x(1-2b_x)}{\cos 2\beta (T_x(1-2b_x) + T_y(1-2b_y))}\right)^2} \quad \text{(Eq. S18)}$$

In addition, the relative amplitude $\mathcal{A}$ of the polarization rotation signal

$$\mathcal{A} = e^{-\sigma n l} \cos 2\beta \frac{T_x(1-2b_x) + T_y(1-2b_y)}{2}$$
$$\times \sqrt{1 - \left(\frac{T_y(1-2b_y) - T_x(1-2b_x)}{\cos 2\beta (T_x(1-2b_x) + T_y(1-2b_y))}\right)^2} \quad \text{(Eq. S19)}$$

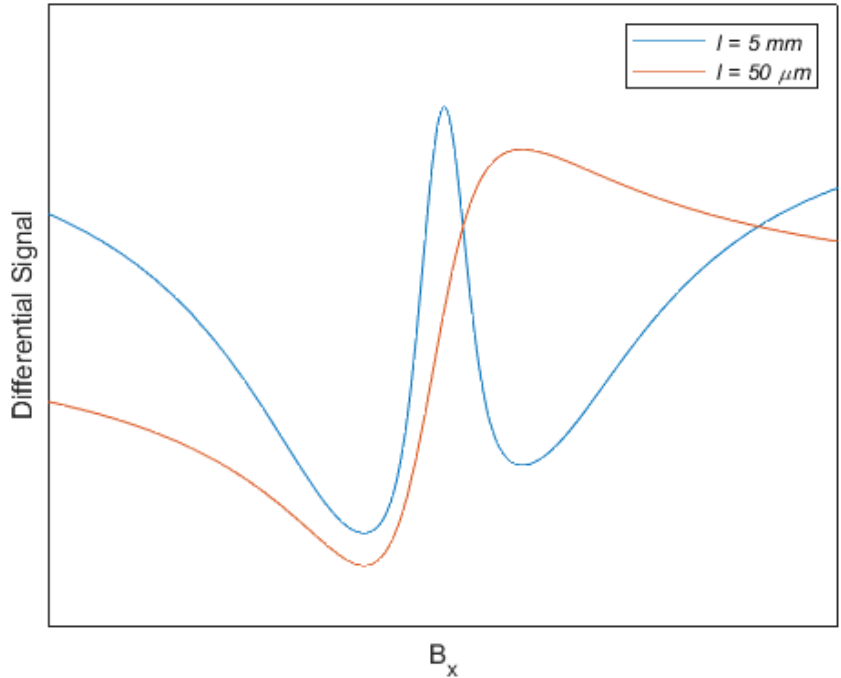

FIG. S1. Simulated differential signal $\mathcal{D}$ due to circular dichroism for a vapor cell with optical path length of 5 $mm$(blue) and 50 $\mu m$(red). The figure is not to scale and only shows the overall shape of the signal.

should be maximized once (Eq. S18) is satisfied.

Using the laser detuning, ellipticity, and linewidth specified in [1], we rewrite the condition in (Eq. S18) in terms of the figure of merit introduced in Section 3 of the main text:

$$FOM = \left| \frac{T_y(1 - 2b_y) - T_x(1 - 2b_x)}{T_x(1 - 2b_x) + T_y(1 - 2b_y)} \right| \ll 0.707 \quad \text{(Eq. S20)}$$

Considering that the $FOM$ is bounded to $\leq 1$, we therefore see that (Eq. S18) can be easily satisfied. In addition, the signal can be further optimized by choosing a different ellipticity $\beta$ and detuning $\nu_0 - \nu$. For a sufficiently small vapor cell such that $\Lambda = n\sigma l P_z \ll 1$, (Eq. S18) is no longer required, and we can instead maximize the signal in the linear $P_z$ regime:

$$\mathcal{D} = \xi E_0^2 e^{-\sigma n l} \cos 2\beta \sin \phi \quad \text{(Eq. S21)}$$

**S4: Effect of laser intensity variation on the pumping rate and the differential signal**

The differential signal in Eq. S21 has a dependence on the polarization rotation angle $\phi$, whose expression has been derived in [1]:

$$\phi = c r_e f n l P_z \frac{\nu_0 - \nu}{(\nu_0 - \nu)^2 + \left(\frac{\Delta \nu}{2}\right)^2} \quad \text{(Eq. S22)}$$

Here $r_e$ is the classical electron radius, $f$ is the oscillator strength of the rubidium D1 transition. The component of $P_z$ at modulation frequency is given by

$$P_z = \frac{2sR\gamma B_x (R + \frac{1}{T_2})^2}{\gamma^2 B_x^2 (R + \frac{1}{T_2})^2 + 1} \quad \text{(Eq. S23)}$$

Here $\gamma$ is the electron gryomagnetic ratio, $Q(P) = 6$ is the nuclear slowdown factor for $^{87}Rb$ in the low polarization limit, $s = i\frac{E \times E^*}{E_0^2} = -\sin 2\beta \hat{z}$ is the photon spin, $T_2$ is the transverse spin relaxation time, $= \sigma \frac{c \epsilon_0 E_0^2}{2h\nu}$ is the optical pumping rate for unpolarized atoms where $\sigma = c r_e f \frac{\frac{\Delta \nu}{2}}{(\nu_0 - \nu)^2 + \left(\frac{\Delta \nu}{2}\right)^2}$.

For a relative intensity variation $\epsilon_{int} = \frac{\Delta I}{I} = \frac{\Delta R}{R}$ (since $R \propto I$), we have a field measurement error of:

$$e(B_x) = \frac{\partial \mathcal{D}}{\partial R} \Delta R \frac{1}{\left|\frac{\partial \mathcal{D}}{\partial B}\right|} \quad \text{(Eq. S24)}$$

For $\epsilon_{int} = 1\%$, Fig.2 shows the field measurement error $e(B_x)$ as a function of $B_x$. We can see that the relationship is mostly linear, indicating a constant relative error of $\frac{e(B_x)}{B_x} = 1.16\%$. We also see in Fig. S2 that this error is almost identical for the metasurface PBS as well as ideal PBS.

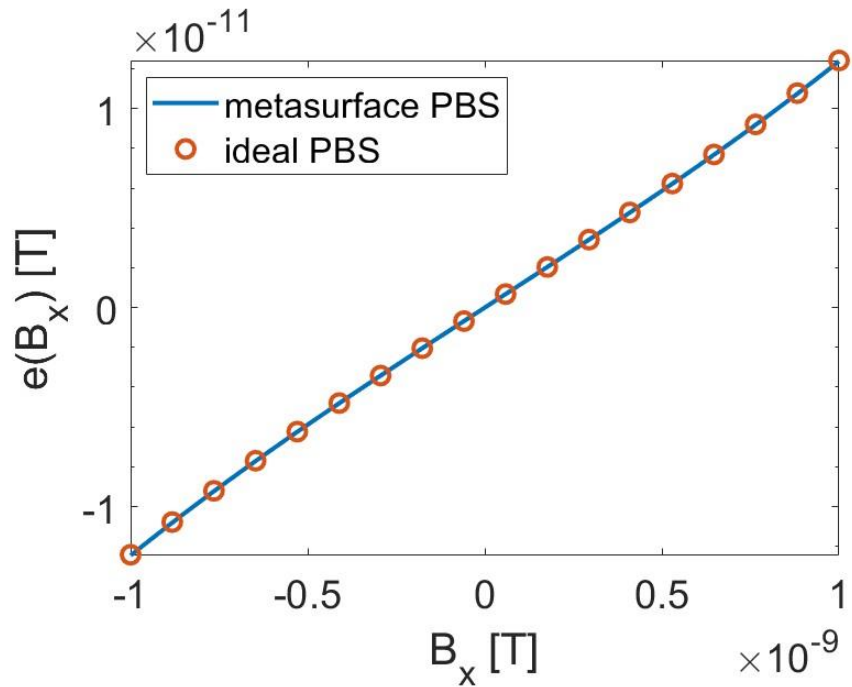

FIG. S2. Simulated magnetic field measurement error due to variation in transmitted light power and optical pumping rate.

References for the Supplementary Information

1. V. Shah and M. V. Romalis, "Spin-exchange relaxation-free magnetometry using elliptically polarized light," Phys. Rev. A **80**(1), 013416 (2009).